\documentstyle[12pt]{article}  
\def\Box{\hbox{$\sqcup$\kern-0.66em\lower0.03ex\hbox{$\sqcap$}}}

\begin{document}
\begin{titlepage}
\begin{flushright}
IFUP--TH 20/98
\end{flushright}
\vskip 1truecm
\begin{center}
\Large\bf
Super Liouville action for Regge surfaces
\footnote {This work is  supported in part
  by M.U.R.S.T.}.
\end{center}

\vskip 1truecm
\begin{center}
{Pietro Menotti}\\ 
{\small\it Dipartimento di Fisica dell'Universit\`a, Pisa 56100, 
Italy and}\\
{\small\it INFN, Sezione di Pisa}\\
{and} \\
{Giuseppe Policastro}\\ 
{\small\it Scuola Normale Superiore, Pisa 56100, 
Italy and}\\
{\small\it INFN, Sezione di Pisa}\\
\end{center}
\vskip .8truecm
\begin{center}
May 1998
\end{center}
\end{titlepage}

\begin{abstract} 
We compute the super Liouville action for a two
dimensional Regge surface by exploiting the invariance of the theory
under the superconformal group for sphere topology and under the
supermodular group for torus topology. For sphere topology and 
torus topology with even spin structures, the action is completely fixed
up to a term which in the continuum limit goes over to a topological
invariant, while the overall normalization of the action can be taken
from perturbation theory.  For the odd spin structure on the torus,
due to the presence of the fermionic supermodulus, the action is fixed
up to a modular invariant quadratic polynomial in the fermionic zero
modes. 
\end{abstract}

\section{Introduction}

Discretized models of field theory serve the purpose of reducing the
infinite number of degrees of freedom to a finite one.
Here we shall be concerned with a discrete approach to two dimensional
supergravity.

In two dimensions on the continuum, the reduction of the functional
integral to the superconformal gauge gives rise to the super Liouville
action \cite{poly2, martinec}. In theories related to gravity several
discretization schemes have been proposed \cite{discrete} one of which
is to approximate a smooth 
manifold by one which is everywhere flat except for a finite number of
$D-2$ dimensional simplices i.e. the Regge model \cite{regge}.

In principle one can think of other schemes of reducing the number of
degrees of freedom to a finite one; e.g. in two dimensional gravity one
could expand the conformal factor, in the case of spherical topology,
in spherical harmonics on the surface of the sphere and keep only a
finite number of modes.

However the Regge scheme has the remarkable advantage that the family
of Regge conformal factors is closed under the invariance groups of
the theory which are $SL(2,C)$ for spherical topology and modular group
for torus topology. This would not occur e.g. by expanding in
spherical harmonics and keeping a finite number of modes because under
a $SL(2,C)$ transformation they would mix with an infinite number of
modes.

In ref.\cite{pmppp} the ordinary non supersymmetric case was
considered; the Liouville action for a Regge surface and also the
measure for the conformal factor was derived and shown that the
resulting theory is exactly invariant under $SL(2,C)$ for sphere
topology. For torus topology the procedure provided also a non formal
explicit proof of the modular invariance of the theory. Moreover the
derived measure is Weyl invariant \cite{dhoker,pmpppsanta} thus
providing a Weyl invariant discretization scheme.
 
The procedure for computing the Liouville action was to exploit the
heat kernel technique similarly to what is done on the continuum. It
was later realized \cite{pm} that the same result (except for a
function which in the continuum limit contributes to a topological
term) can be more easily obtained by exploiting the invariance of the
action under the $SL(2,C)$ group, for sphere topology and under the
modular group for torus topology.  In this paper we extend such a
treatment to the supersymmetric case. Here the role is played by the
superconformal and by the supermodular groups acting on the
superconformal factor.

Obviously one could also follow the standard heat kernel procedure as
it was done in \cite{pmppp} for the non supersymmetric case i.e. by
computing the short time behavior of the heat kernel on a singular
Riemann surface and by choosing the correct self adjoint extension of
the Lichnerowicz-De Rahm operator by exploiting the Riemann-Roch
theorem. The procedure in the supersymmetric case is further
complicated by the fact that the superlaplacian is not a definite
positive operator and thus in order to apply the heat kernel procedure
one has to compute the heat kernel of the square of the operator and
then take the square root of the result.

Here we shall follow the simpler procedure of ref.\cite{pm}. We shall
see that for sphere topology and for the even spin structures for
torus topology the action is fixed up to a function of the conical
defects which in the continuum limit goes over to a topological
invariant. The overall normalization of the action can be taken from
perturbation theory. For the odd spin structure on the torus, due to
the presence of the fermionic supermodulus and the associated
fermionic zero modes, group theory determines the action up to a
polynomial quadratic in the amplitudes of the two fermionic zero
modes. Modular invariance imposes certain restrictions on the
coefficients of such polynomial which however are not sufficient to
determine them completely. Thus it appears that for the odd spin
structure case a closer appeal the structure of the the heat kernel
derivation is necessary to fix the action completely.

The computation of the integration measure for the superconformal
factor which is the remaining ingredient in the discretized functional
integral, is left for an other paper.
  
The two dimensional Regge surface in \cite{pmppp} was described by a
conformal factor given in terms of the positions and the strengths of
the singularities to which, for torus topology the Teichm\"uller
parameters have to be added. We recall that such a description is
completely equivalent to the usual one in terms of triangulations but
in two dimensions the  use of complex coordinates appears
more powerful. Similarly in the supersymmetric case, exploiting a well
known result by Howe \cite{howe}, we shall describe the supergeometry
in terms of a superconformal factor supplemented in the case of torus
topology, by the supermoduli. The idea of using the complex plane and
supercomplex plane to describe a Regge geometry is due to Foerster
\cite{foer} where he also makes an heuristic guess of the action and
of the integrations measure; such guesses however do not agree with
the exact results of \cite{pmppp}.

The supersymmetric approach that we shall describe in the following
allows also to introduce spinning particle in a natural way on a Regge
surface.

\section{Sphere topology}

As usual we shall describe the supersphere by a single chart given by
the complex superplane ${\bf z}=(z,\theta)$ with $z=x+iy$, completed by
the point at 
infinity. It was proven in \cite{howe} that any two dimensional
supergeometry is locally superconformally flat. As for the sphere
topology there are no Teichm\"uller parameters, the
supergeometry of the supersphere can be given in terms of a super Weyl
transformation applied to the flat background described by the
superzweibein 
\begin{equation} 
\label{flat}
\hat E_M^{~~A} =\left(  \matrix {\delta_m^{~~a} &
0  \cr \frac{1}{2} (\gamma^a)_\mu^{~~\beta}\theta_\beta &
\delta_\mu^{~~\alpha} \\ } 
  \right) = \left(  \matrix {1 & 0 & 0 & 0 \cr
0 & 1 & 0 & 0 \cr \theta & 0 & 1 & 0 \cr 0 & -\bar\theta & 0 & 1 }
  \right)
\end{equation}
and zero connection.
The indices $M = (m,\mu)$ are Einstein indices; $m$ runs over the
values $z,\bar{z}$ while $\mu$ runs over $\theta,\bar{\theta}$. The
indices $A = (a,\alpha)$ are Lorentz internal indices; $a$ runs over
the values $u,\bar{u}$ and $\alpha$ over the values +,$-$.
The new zweibein resulting from the super Weyl transformations is
given by \cite{howe}
\begin{equation} \label{Weyl}
E_M^{~~a} = e^\Sigma\hat E_M^{~~a};~~~~E_M^{~~\alpha} =
e^{\Sigma/2}\hat E_M^{~~\alpha}~-2~\hat
E_M^{~~b}(\gamma_b)^{\alpha\beta} \hat D_{\beta}e^{\Sigma/2}
\end{equation}
being
\begin{equation} 
{\hat {\cal D}}_+= \hat E_+^{~M}\partial_M = {\partial\over \partial
\theta} - \theta 
{\partial\over \partial z};~~~~\hat{\cal D}_-=\hat E_-^{~M}\partial_M=
{\partial\over 
\partial  \bar \theta} + \bar \theta
{\partial\over \partial \bar z}
\end{equation} 
where $E_A^{~M}$ denotes the inverse superzweibein. The superzweibein
(\ref{Weyl}) satisfies the torsion constraints
\cite{howe,wz,dhph} 
for any $\Sigma$.

This description of the supergeometry is invariant under the
superconformal group given by \cite{dhph}
\begin{equation}
\label{superconformal}
z'= {az+b+\alpha \theta\over cz +d + \beta \theta};~~~~\theta'= {\gamma
z+\delta +A \theta\over cz +d + \beta \theta} 
\end{equation}
with the restrictions
\begin{equation}
\begin{array}{cc}
ad-bc+\gamma\delta =1; & a\beta-c\alpha+A\gamma=0 \\
b\beta-d\alpha+A\delta=0; & A^2-2\alpha\beta=1
\end{array}
\end{equation}
On the continuum the super Liouville action is given by
\cite{poly2,martinec,dhph}
\begin{equation}
\label{sliouville} 
S_{sL}(\Sigma) =
-\frac{10}{8\pi}\int d^2{\bf z} \tilde E
( \tilde D_+ \Sigma \tilde D_-\Sigma + \tilde {\cal R}_{+-} \Sigma)
\end{equation}
where $\tilde E$ is the superdeterminant of the background
superzweibein.  Taking into account
that \cite{dhph} 
\begin{equation}
\label{curvature}
{\cal R}_{+-} =e^{-\Sigma}(\tilde{\cal R}_{+-} -2
\tilde D_+\tilde D_-\Sigma)\equiv e^{-\Sigma}(\tilde{\cal R}_{+-} -2
\tilde{\Box}^{(-)}\Sigma)
\end{equation}
being ${\cal R}_{+-}$ the supercurvature, with an integration
by parts eq.(\ref{sliouville}) can be reduced to Polyakov's non local
covariant form
\begin{eqnarray}
\label{sLaction}
S_{sL} =  \frac{10}{32\pi}\left[ \int d^2{\bf z} (E
{\cal R}_{+-})({\bf z}) G({\bf z}, {\bf z}') d^2{\bf z}'  ( E
{\cal R}_{+-})({\bf z}') \right. 
\nonumber \\
\left.  - 4 \ln(\frac{A}{A_0})\int d^2{\bf z} E {\cal R}_{+-} \right] 
\end{eqnarray}
where $A$ is the area and $A_0$ is the reference area \cite{pmppp}.
The origin of the last term is due to the contribution of the
zero modes as stressed in \cite{duffdowker}.
$G$ is the Green function of the operator $\Box^{(-)}$ i.e.
\begin{equation}
\ \Box^{(-)} G({\bf z}, {\bf z}') =
\delta^2(z-z') (\bar\theta - \bar\theta') (\theta - \theta')
\frac{1}{E({\bf z}')}.
\end{equation}
As $\Box^{(-)}=e^{-\Sigma}\hat{\Box}^{(-)}$ we have that $G({\bf z},
{\bf z}')=\hat G({\bf z},{\bf z}')$  defined by
\begin{equation}
\label{greeneq}
\hat{\Box}^{(-)}\hat G ({\bf z},{\bf z}') = \delta^2(z-z')(\bar\theta -
\bar\theta ')(\theta- \theta').
\end{equation} 
The solution of eq.(\ref{greeneq}) is \cite{dhph}
\begin{equation}
\hat G({\bf z}, {\bf z}') = \frac{1}{\pi}
\ln [(z-z' +\theta\theta')(\bar z-\bar z'
-\bar\theta\bar\theta')+\varepsilon^2] 
\end{equation}  
being $z-z' +\theta\theta'$ the superinvariant displacement. In the
following the $\varepsilon^2$ will be understood.
It follows that the superconformal factor $\Sigma$ describing a super
Regge two dimensional surface with the topology of the sphere is given
by
\begin{equation}
\label{conffac}
\Sigma ({\bf z}) = \sum_i (\alpha_i-1)\ln [(z-z_i
+\theta\theta_i)(\bar z-\bar z_i-\bar\theta\bar\theta_i)]+\lambda_0 
\end{equation}
being $1-\alpha_i =\delta_i$ the
conical defects at the points $(z_i,\theta_i)$. The $\alpha_i$ are
constrained by the Gauss-Bonnet theorem
\begin{equation}
\int d^2 z d\theta\, d\bar\theta\, {\cal R}_{+-}= \pi \chi({\cal M}),
\end{equation}
being $\chi({\cal M})$ the Euler characteristic of the manifold. 
Substituting eq.(\ref{conffac}) into eq.(\ref{curvature}) one obtains
\begin{equation}
\label{gaussbonnet}
2\pi = - 2\pi \sum_i \int d^2 z \delta(z-z_i)(\alpha_i-1) = 2\pi
\sum_i\delta_i \, . 
\end{equation}
We note that this is at variance with the
ordinary non supersymmetric case where $\sum_i \delta_i =2$; it is due
to the fact that $\Sigma$ is related to the superzweibein instead of
to the metric.
We recall that under a superconformal transformation the displacement
${\bf z}_{ij} = z_i - z_j + \theta_i\theta_j$ behave as
\begin{equation}
{\bf z}_{ij}' = z_i' - z_j' + \theta_i'\theta_j'  = \frac{{\bf z}_{ij}
}{(cz_i+d+\beta\theta_i)(cz_j+d+\beta\theta_j)}\,. 
\end{equation}
The corresponding transformation of the superconformal factor is as
follows
\begin{eqnarray}
\Sigma'(z',\theta';z_i,\theta_i,\alpha_i,\lambda_0)  = 
\Sigma (z(z',\theta'),\theta(z',\theta');
z_i,\theta_i,\alpha_i,\lambda_0)   + \nonumber \\ 
+ \ln \left|{\rm sdet} \frac{\partial (z,\bar
z,\theta,\bar\theta)}{\partial (z',\bar z',\theta',\bar\theta')}\right| =
\nonumber  \\
% = \lambda_0 + \sum_i (\alpha_i-1) \ln (z-z_i+\theta\theta_i)(\bar
%z-\bar z_i-\bar\theta \bar\theta_i) \nonumber \\
%+ \ln (cz+d+\beta\theta)(\bar c \bar z+\bar d-\bar\beta\bar\theta) =
\nonumber \\ 
\sum_i (\alpha_i-1) \ln (z'-z_i'+\theta'\theta_i')(\bar
z'-\bar z_i'-\bar\theta' \bar\theta_i')\nonumber + \lambda_0\\
+ \sum_i (\alpha_i-1) \ln (cz_i +d+\beta\theta_i)(\bar c
\bar z_i +\bar d-\bar\beta\bar\theta_i)
\end{eqnarray}
where we have taken into account the constraint $\sum_i (1-\alpha_i) =
1$.
Thus we have 
\begin{eqnarray}
\Sigma'(z',\theta';z_i,\theta_i,\alpha_i,\lambda_0)  = 
\Sigma(z',\theta';z_i',\theta_i',\alpha_i',\lambda_0')
\end{eqnarray}
with
\begin{equation}
z_i' = \frac{az_i+b+\alpha\theta_i}{cz_i+d+\beta\theta_i} ; ~~~~ 
\theta_i' = \frac{\gamma z_i + \delta
+A\theta_i}{cz_i+d+\beta\theta_i} 
\end{equation}
\begin{equation}
\begin{array}{cc}
\lambda_0' = \lambda_0 + \sum_i (\alpha_i-1) \ln (cz_i
+d+\beta\theta_i) (\bar c \bar z_i +\bar d-\bar\beta\bar\theta_i);
\nonumber \\ 
\alpha_i' = \alpha_i.
\end{array}
\end{equation}
The condition $\sum_i \delta_i = 1$ plays a crucial role in the above
transformation which shows that the family of Regge superconformal
factors is closed under the superconformal group.

The discrete transcription of the nonlocal action (\ref{sLaction})
is 
\begin{equation} \label{discrete}
S_{sL} = \frac{1}{2} \sum_{i,j} K_{ij} [\alpha] \ln [(z_i-z_j  +
\theta_i \theta_j)(\bar z_i-\bar z_j  -\bar\theta_i \bar\theta_j)] +
B(\lambda_0, \alpha)
\end{equation}
with $K_{ij} = K_{ji}$ and $K_{ii}=0$.
Under a superconformal transformation the action goes over to
\begin{eqnarray}
S_{sL} \rightarrow S_{sL} - \sum_{ij} K_{ij} [\alpha] \ln
(cz_i+d+\beta\theta_i) \nonumber \\
- \sum_{ij} K_{ij} [\alpha] \ln (\bar c \bar z_i+\bar d-\bar\beta\bar\theta_i)
+ B(\lambda_0',\alpha) -B(\lambda_0,\alpha)
\end{eqnarray}
and thus we must have
\begin{eqnarray}
- \sum_{ij}K_{ij}[\alpha]  \ln (cz_i+d+\beta\theta_i) 
- \sum_{ij} K_{ij} [\alpha] \ln (\bar c \bar z_j+\bar d-\bar\beta\bar\theta_j
)+ B(\lambda_0',\alpha) = B(\lambda_0,\alpha).
\end{eqnarray}
In order to find the structure of $B$ let us consider first the
transformation eq.(\ref{superconformal})
\begin{equation}
a=\frac{1}{d}=k;~~~~A=1
\end{equation}
and all other parameters equal to zero.
Then we have
\begin{equation}
2 \ln k\sum_{ij} K_{ij} [\alpha]  + B(\lambda_0',\alpha) =
B(\lambda_0, \alpha)
\end{equation}
with $\lambda_0'=\lambda_0 +2 \ln k$, which tells us that $B$ is a
linear function of $\lambda_0$ i.e.
\begin{equation}
B(\lambda_0,\alpha) = - \lambda_0 \sum_{ij} K_{ij} [\alpha] +
F[\alpha]
\end{equation}
so reaching the structure
\begin{equation}
S_{sL} = \frac{1}{2}\sum_{ij}  K_{ij} [\alpha] \ln ({\bf z}_{ij} \bar{\bf
z}_{ij}) - \lambda_0  \sum_{ij} K_{ij} [\alpha] + F[\alpha].
\end{equation}
Imposing invariance under the general superconformal
transformation we have
\begin{eqnarray}
0= -  \sum_{ij} K_{ij} [\alpha] \ln  (cz_i +d+\beta\theta_i)(\bar c
\bar z_i +\bar d-\bar\beta\bar\theta_i) - \nonumber \\\sum_i (\alpha_i-1) \ln
(cz_i +d+\beta\theta_i)(\bar c \bar z_i +\bar d-\bar\beta\bar\theta_i)
\sum_{mn} K_{mn} [\alpha]
\end{eqnarray} 
which gives, due to the arbitrariness of $z_i$ and $\theta_i$,
\begin{equation}
0 = \sum_j K_{ij} [\alpha] + (\alpha_i-1) \sum_{mn} K_{mn} [\alpha].
\end{equation}
Defining $K_{ij}[\alpha] = (1-\alpha_i)(1-\alpha_j) h_{ij}[\alpha]$
the above equation becomes
\begin{equation} \label{hfactor}
\sum_{j\neq i} (1-\alpha_j) h_{ij}[\alpha] = \sum_{m,n\neq m}
(1-\alpha_m)(1-\alpha_n) h_{mn}[\alpha].
\end{equation}
We assume that $h_{mn}[\alpha]$ will depend only on
$\alpha_n$, $\alpha_m$, i.e.
\begin{equation}
h_{mn}[\alpha] = h(\alpha_m,\alpha_n).
\end{equation}
Let us choose the $\alpha_i ~~(i=1,N)$ as follows: $\alpha_1$ and
$\alpha_2$ free and for $i \geq 3 ~~~\alpha_i = \bar \alpha$ with
\begin{equation}
1-\bar\alpha = \frac{\alpha_1 + \alpha_2 -1}{N-2}.
\end{equation}
Substituting into (\ref{hfactor}) we obtain with $i=1$
\begin{eqnarray} \label{N}
(1-\alpha_2)(2\alpha_1-1)h(\alpha_1,\alpha_2) +
(\alpha_1+\alpha_2-1)(2\alpha_1-1)h(\alpha_1,\bar\alpha) = \nonumber \\2
(1-\alpha_2)(\alpha_1+\alpha_2-1) h(\alpha_2,\bar\alpha) +
\frac{(N-2)(N-3)}{(N-2)^2}(\alpha_1+\alpha_2-1)^2h(\bar\alpha,\bar\alpha).
\end{eqnarray}
The above equation has to hold for any $N$, and in the limit $N
\rightarrow \infty$ it reduces to
\begin{eqnarray} \label{infty}
(2\alpha_1-1)(1-\alpha_2)h(\alpha_1,\alpha_2) +
(\alpha_1+\alpha_2-1)(2\alpha_1-1)h(\alpha_1,1) = \nonumber \\
2(1-\alpha_2)(\alpha_1+\alpha_2-1)h(\alpha_2,1) +
(\alpha_1+\alpha_2-1)^2 h(1,1).
\end{eqnarray}
Setting $\alpha_2=1$ we obtain
\begin{eqnarray}
\alpha_1(2\alpha_1-1) h(\alpha_1,1) = \alpha_1^2 h(1,1)~~~~ \textrm{i.e.}
\nonumber \\
h(\alpha_1,1) = \frac{\alpha_1}{2\alpha_1-1}h(1,1)
\end{eqnarray}
and substituting into (\ref{infty}) we finally obtain
\begin{equation}
h(\alpha_1,\alpha_2)=h(1,1) \frac{1}{2} \left( \frac{1}{2\alpha_1-1} +
\frac{1}{2\alpha_2-1} \right).
\end{equation}
We notice that the derived $h(\alpha_1,\alpha_2)$ satisfy exactly
eqn. (\ref{N}) for any $N$ and also eq.(\ref{hfactor}).
In conclusion the action takes the form 
\begin{equation}
S_{sL} = {\rm const} \left[  \sum_{i,j\neq i}
\frac{(1-\alpha_i)(1-\alpha_j)}{(2\alpha_i-1)} \ln ({\bf z}_{ij}
{\bar {\bf z}}_{ij}) + \frac{\lambda_0}{2} \sum_i \left(
2\alpha_i-1 - \frac{1}{2\alpha_i-1}  \right) + F[\alpha] \right].
\end{equation}
Group theory alone cannot fix the proportionality constant and the
function $F[\alpha]$. The latter, as happens in the usual non
supersymmetric case, represents the analogue of the renormalized
electrostatic self energies of the point charges (singular curvatures)
and from the structure of the heat kernel derivation of
eq.(\ref{discrete}) we know it will have the form $F[\alpha] = \sum_i
f(\alpha_i)$.
The explicit form of $f(\alpha)$ can be obtained only from the full
heat kernel derivation \cite{martinec}, as in the usual case. However, in the
continuum limit it becomes
\begin{equation}
\sum_i f(\alpha_i) = \sum_i f(1) - f'(1) \sum_i \delta_i = N f(1) -
f'(1) 
\end{equation}
which, due to (\ref{gaussbonnet}) is a constant term of topological
nature.
The proportionality constant can be borrowed from perturbation theory
and has the value $5/4$.
Summing up, for sphere topology we have 
\begin{equation} \label{sphereaction}
S_{sL} = \frac{5}{4} \left( \sum_{i,j \neq i}
\frac{(1-\alpha_i)(1-\alpha_j)}{2\alpha_i-1} \ln ({\bf z}_{ij}
{\bar {\bf z}}_{ij}) + \frac{\lambda_0}{2} \sum_i \left( 2\alpha_i-1 -
\frac{1}{2\alpha_i-1} \right) + \sum_i f(\alpha_i) \right)
\end{equation}
This expression, for $\alpha_i \simeq 1$ and for a dense set of ${\bf
z_i}$ goes over to the continuum result (\ref{sLaction}). 

\section{Torus topology}

The supertorus can be defined \cite{crrb} as the quotient of the
complex superplane ${\bf C}^{(1,1)}$ with respect to an abelian group $G$
(the fundamental group of the torus) with two generators $g_1,g_2$
which act on  ${\bf  C}^{(1,1)}$ in a properly discontinuous manner,
leaving  invariant the metric element $(dz+d\theta\,\theta)(d\bar z
-d\bar\theta\,\bar\theta)$. We will denote $g$ by $(a,b,\alpha)$, i.e. 
$(a,b,\alpha)(z,\theta)=(z+ab+\alpha\theta,a\theta+a\alpha)$.
The conditions imposed on $g_1$ and $g_2$ imply that $a_1$ and $a_2$
are equal to $\pm 1$. We must distinguish between two cases\\
1. {\it Even spin structures}

These are given by $a_1$ and $a_2$ not both equal to 1. It is known
\cite{crrb} that by means of a conjugation with an element of $G$ and with
a uniform rescaling $z\rightarrow k^2 z,~ \theta\rightarrow k\theta$,
the generators can be reduced to the canonical form 

$g_1=(a_1,1,0) \\
\indent  g_2=(a_2,\tau,0)$ \\
where $\tau$ is a bosonic modulus. \\
2. {\it Odd spin structure}

This is given by $a_1=a_2=1$. By means of a conjugation with a uniform
rescaling we can reduce the generators to the form

$g_1=(1,1,\alpha) \\
\indent g_2=(1,\tau,c\alpha). $ \\
If furthermore one conjugates with respect to the element of the
superconformal group $f(z,\theta) = (z(1+\alpha\theta),\theta+\alpha
z)$ one obtains \cite{crrb}

$g_1=(1,1,0) \\
\indent g_2 = (1,\tau,\chi)$ \\
where $\chi=(c-\tau)\alpha$ is the fermionic supermodulus which is
absent in the even spin structures.
The superzweibein is not left unchanged under such a transformation,
instead it goes over to
\begin{equation}
\left( \matrix{1+2\alpha\theta' & \alpha \cr \theta' & 1} \right)
\end{equation}
for the $(z,\theta)$ sector, and similarly for the $(\bar z,
\bar\theta)$ sector. However, such a zweibein can be obtained from
$\hat E$ by means of the super Weyl transformation generated by
$e^{\Sigma} = (1+\alpha\theta')(1-\bar\alpha \bar\theta')$,
supplemented by the internal U(1) transformation given by the rotation
$\displaystyle{\frac{1+\bar\alpha \bar \theta'}{1-\alpha\theta'}}$. 
As the superdeterminant of the last transformation is 1, everything
reduces to a change in the superconformal factor.
 
In case 1, i.e. even spin structures, the modular group is given by
the usual modular transformations
\begin{eqnarray}
z'=\frac{z}{c\tau +d} & & \theta'=\frac{\theta}{(c\tau +d)^{1/2}}
\nonumber \\
\tau'=\frac{a\tau+b}{c\tau +d} & & \textrm{with}~ ad-bc=1,~ a,b,c,d \in
{\bf Z}   
\end{eqnarray}
which are generated by the two transformations 
\begin{equation}
\tau'=\tau +1 ~~~\textrm{and} ~~\tau' = -\frac{1}{\tau}.
\end{equation}
Writing $z=x+\tau y$, $\bar z=x+\bar \tau y$ the fundamental region in
$(x,y)$ is $[0,1] \times [0,1]$.

In case 2, i.e. odd spin structure, we have the supermodular
transformations  given by 
\begin{eqnarray} \label{supermodular1}
z'=\frac{z}{c\tau +d + c\chi \theta}; & & \theta'=\frac{\theta}{(c\tau
+d)^{1/2}}-\frac{c \chi z}{(c\tau +d)^{3/2}}; 
\end{eqnarray}
\begin{eqnarray} \label{supermodular2}
\tau'= \frac{a \tau +b}{c \tau  +d}; & & \chi'=\frac{\chi}{(c\tau
+d)^{3/2}}
\end{eqnarray}
whose generators are
\begin{eqnarray}
(\tau',\chi') & = & (\tau +1,\chi) \nonumber \\
(\tau',\chi') & = & (-\frac{1}{\tau},\frac{\chi}{\tau ^{3/2}}).
\end{eqnarray}
We recall moreover that the two elements 
$g=(1,\tau,\chi)$ and $(1,\tau, e^{i\phi}\chi)$ with
$\phi=n\frac{\pi}{2}$ are equivalent under conjugation \cite{crrb}.

On a torus equipped with an even spin structure
$(a, b)$ with $a_1=(-1)^a,~a_2=(-1)^b$ the flat 
superzweibein is still given by eq. (\ref{flat}).
The Green function $\hat G_{ab} (\bf z,\bf z'; \tau)$ satisfies  
\begin{equation}
\int d^2 {\bf z}' \hat{\Box}^{(-)} \hat G_{ab}   ({\bf
z},{\bf z}'; \tau) 
f_{ab} ({\bf z}') = f_{ab} ({\bf z})
\end{equation}
for any $f_{ab}({\bf z})$ belonging to the $(a,b)$ spin
structure and orthogonal to the zero modes of $\hat{\Box}^{(-)}$
i.e. such that $\int dzd\bar z d\theta d\bar \theta f_{ab} =0$; in
fact even if the supermetric $( f_{ab} , f_{ab}) = \int d^2{\bf z}
\hat E \bar f_{ab} f_{ab} $ 
is not definite positive, one can explicitly  check that
${\rm Range}({\Box}^{(-)})= ({\rm Ker}({\Box}^{(-)}))^\bot$. Such a
Green function is given by
\cite{dhph,mumford,tanii} 
\begin{equation}
G_{ab}({\bf z},{\bf z}',\tau) = G_0(z-z',\tau) +
\left[ \theta \theta' 
S_{ab}(z-z') + c.c. \right] \qquad (\textrm{for}~ z \neq z')
\end{equation}
where \cite{mumford}
\begin{equation}
G_0(z-z',\tau) = \frac{1}{\pi} \ln \left|
\frac{\vartheta_{11}(z-z',\tau)}{\eta(\tau)} \right| ^2 + 
\frac {i (z - \bar z - z' + \bar z')^2}{\tau -\bar \tau}
\end{equation}
and 
\begin{equation}
S_{ab}(z-z') = \frac{1}{\pi}~
\frac{\vartheta_{11}'(0,\tau)}{\vartheta_{ab}(0,\tau)}~
\frac{\vartheta_{ab}(z-z',\tau)}{\vartheta_{11}(z-z',\tau)}. 
\end{equation}
The transformation properties of the Green functions under modular
transformations are the following

for $(z,\theta) \rightarrow (z,\theta)$ and $\tau \rightarrow \tau
+1$, $ \qquad G_{ab} \rightarrow G_{a,a+b+1}$; 

for $(z,\theta) \rightarrow
(\frac{z}{\tau},\frac{\theta}{\tau^{1/2}})$ and $\tau \rightarrow
-\frac{1}{\tau}$, $ \qquad G_{ab} \rightarrow G_{ba}; \\ $
(with $a$ and $b$ defined modulo $2$) that is, the even spin
structures $(0,0), (0,1)$ and $(1,0)$ undergo a permutation.   

We come now to the odd spin structure. For clearness sake we shall
denote by $ \hat z, \hat \theta$ the original variables $z,\theta$
which appear in eq.(\ref{supermodular1},\ref{supermodular2}) and are
associated to the zweibein $\hat E$ of eq.(\ref{flat}). One can
perform the change of variables
\begin{equation} \label{hatvar}
\hat z= z + \zeta (z- \bar z) \theta ; \qquad 
\hat \theta = \theta + \zeta (z-\bar z)
\end{equation}
where $\zeta = \frac{\chi}{\tau -\bar\tau}$ ; the fundamental region
in $z,\theta$ takes a particularly simple form, i.e. the product of
$\theta$ by an ordinary torus of modulus $\tau$.
The superzweibein $\hat E$ is transformed into 
\begin{equation} \label{oddzweibein}
\tilde E_M^{~A} = \left( \matrix {1+2\zeta\theta & 2\bar\zeta
\bar\theta & \zeta & -\bar\zeta \cr 
-2\zeta\theta & 1-2\bar\zeta\bar\theta & -\zeta & \bar\zeta \cr
\theta & 0 & 1 & 0 \cr 0 & -\bar\theta & 0 & 1 \cr } \right).
\end{equation}
$\tilde E_M^{~A} $ is invariant under the usual translations and the
two super Killing vectors
\begin{equation}
\begin{array}{ll}
z'=z + \varepsilon \zeta \theta; & \theta' = \theta + \varepsilon \zeta
\\ \bar z' = \bar z - \bar \varepsilon \bar\zeta \bar \theta; &
\bar \theta' = \bar \theta +\bar\varepsilon \bar\zeta .
\end{array}
\end{equation} 
The superinvariant derivatives are 
\begin{eqnarray} \label{oddderiv}
\tilde {\cal D_+} = (1+\theta\zeta+\theta\bar\theta \zeta \bar\zeta)
\partial_{\theta} - \theta \bar\zeta \partial_{\bar\theta} - \theta
\partial_z -\theta\bar\theta\bar\zeta \partial_{\bar z}; \nonumber \\
\tilde {\cal D_-} = (1-\bar\theta\bar\zeta + \theta \bar\theta
\zeta\bar\zeta ) \partial_{\bar\theta} + \bar\theta \zeta
\partial_\theta + \theta \bar\theta \zeta \partial_z + \bar\theta
\partial_{\bar z} .
\end{eqnarray}
The Green function of $\tilde {\Box}^{(-)} $ for the odd spin
structure $G_{++}({\bf z}, {\bf z}'; \tau,\chi)$ satisfies  
\begin{equation} \int d^2{\bf z}' \tilde E'
 \tilde{\Box}^{(-)} G_{++}({\bf z},{\bf z}'; \tau,\chi) f_{++} ({\bf
z}') = f_{++} ({\bf z})  
\end{equation}
for any  $f_{++} = f_0 + \theta \phi -\bar
\theta \bar \phi + \theta \bar \theta f_A$ belonging to the odd spin structure
with 
\begin{equation}
\begin{array}{l}
\int d^2z (\phi - \zeta f_0) = {\rm const}~ \zeta \bar\zeta; \nonumber \\
\int d^2z (\bar \phi - \bar \zeta f_0) = {\rm const}~ \zeta \bar\zeta;
\nonumber \\
\int d^2z (f_A -2 \zeta \bar\zeta f_0) =0, 
\end{array}
\end{equation}
i.e. orthogonal to the kernel of $\tilde{\Box}^{(-)}$.
In fact, as it happens for the even spin structures, one can
explicitly check that ${\rm Range}(\tilde{\Box}^{(-)})= ({\rm
Ker}(\tilde{\Box}^{(-)}))^\bot$.
$G_{++}$ is given by a supersymmetric generalization of the function
$G$ of the usual case, i.e. \cite{dhph}
\begin{equation} \label{oddgreenfunc}
\begin{array}{l}
G_{++} ({\bf z},{\bf z}';\tau,\chi) = \nonumber \\
=  \displaystyle{
\frac{1}{\pi} 
\left[  \ln \left|
\frac{\vartheta_{11}(\hat z -\hat z'+\hat\theta \hat\theta',\tilde
\tau)}{\eta(\tilde \tau)} \right| ^2 + \frac{i\pi}{\tilde \tau -
\bar{\tilde \tau}} 
(\hat z-\hat z'+\hat \theta \hat\theta' -\bar {\hat z} + \bar{ \hat
z}' +\bar{\hat\theta} \bar{\hat\theta}')^2 
\right] }
\end{array}
\end{equation}
where $\tilde \tau = \tau + \chi (\hat\theta +\hat\theta')$ and $\hat
z,\hat \theta$ are given by eq.(\ref{hatvar}). It is easily
checked that $G_{++}$, in addition of possessing the two ordinary
Killing vectors $\hat z \rightarrow \hat z + \varepsilon$ and the
two  super Killing 
vectors $\hat z \rightarrow \hat z+\varepsilon\chi\hat \theta, ~\hat
\theta \rightarrow 
\hat \theta+ \varepsilon \chi$, is invariant under the supermodular
transformations eq.(\ref{supermodular1},\ref{supermodular2}) in the
variables $\hat z,~\hat\theta$.

For even spin structures the Regge superconformal factor is given by 
\begin{equation}
\Sigma({\bf z}, {\bf z_i}, \alpha_i,\lambda_0,\tau) =
\sum_i (\alpha_i-1) \pi G_{ab} ({\bf z}, {\bf z_i}, \tau) + \lambda_0
\end{equation}
where $\lambda_0$ is the unique zero mode of $\hat {\Box} ^{(-)}$ in
the space of even spin structure functions.
The Gauss-Bonnet theorem now imposes $\sum_i (1-\alpha_i)=0$.
The discrete transcription of the action (\ref{sLaction}) is given by
\begin{equation}
S_{sL} = {\rm const} \left[ \frac{1}{2} \sum_{ij} K_{ij} [\alpha] \pi G_{ab}
({\bf z_i},{\bf z_j}; \tau) + B(\lambda_0,\alpha,\tau) \right] . 
\end{equation}
We shall now exploit the fact, explicitly verified in the non
supersymmetric case and which follows from the nature of the heat
kernel derivation, that the singular behavior of the action at short
distances is independent of the topology.
Thus we find
\begin{equation}
K_{ij}[\alpha]
=(1-\alpha_i)(1-\alpha_j)(\frac{1}{2\alpha_i-1}+\frac{1}{2\alpha_j-1}).
\end{equation}

Next we impose the invariance of the action under modular
transformations. The superconformal factor
transforms as
\begin{equation}
\Sigma'({\bf z'})=\Sigma({\bf z}) + \ln \left| \tau c +d \right|
\end{equation}
i.e.
\begin{equation}
\Sigma'(z',\theta',z_i,\theta_i,\alpha_i,\lambda_0,\tau) = 
\Sigma(z',\theta',z_i',\theta_i',\alpha_i',\lambda_0',\tau')
\end{equation}
with 
\begin{eqnarray}
z_i'=\frac{z_i}{c\tau +d}; &\qquad
\theta_i'=\displaystyle{\frac{\theta_i}{(c\tau 
+d)^{1/2}}}; &\qquad \alpha_i'=\alpha_i; \nonumber \\
&\lambda_0'=\lambda_0 + \ln \left| c \tau +d \right|; &
\tau'=\frac{a\tau+b}{c\tau+d}. 
\end{eqnarray}
Thus $e^{\lambda_0}$ transforms as the modulus of a modular form of weight
1.
Posing
\begin{equation}
B(\lambda_0,\alpha,\tau)=C(\lambda_0 - \ln \left| 2\pi \eta^2(\tau) \right|
,\alpha,\tau) 
\end{equation}
we have that 
\begin{equation}
C(x,\alpha,\tau)=C(x,\alpha, \frac{a \tau+b}{c \tau+d}).
\end{equation}
To further specialize the structure of $C$ we proceed as in
ref.\cite{pm}
considering the transformation $\lambda_0 \rightarrow \lambda_0-\ln
k^2,~ z_i\rightarrow k^2 z_i,~\theta_i \rightarrow k\theta_i$. Taking into
account that $\sum_i (1-\alpha_i) = 0$ we find
\begin{equation}
B = -2 (\lambda_0 -\ln \left| 2\pi\eta^2(\tau) \right|) \sum_i
\frac{(1-\alpha_i)^2}{2\alpha_i-1} + F[\alpha,\tau].
\end{equation}

The only requirement on $F[\alpha,\tau]$ imposed by group theoretical
considerations is to be a modular invariant function of
$\tau$. However, if we limit ourselves, as in ref. \cite{pm} , to
the realm of modular functions, the only choice which is free of
singularities in the upper half plane, infinity included, is to take
$F$ independent of $\tau$ \cite{serre}. 
This also follows more directly from the nature of the heat kernel
derivation according to which
$F[\alpha]$ has the structure $\sum_i f(\alpha_i)$ . 
In fact from inspection one sees that such a term arises form
the ``direct'' terms \cite{aur} in the variation of the logarithm of
the determinant, which depend only on the short time behavior of the
heat kernel in the proximity of $z_i$ and thus only on the
$\alpha_i$ and are independent of  the topology and the moduli of the
surface. 
Summing up, for the even spin structures we obtain the action
\begin{eqnarray}
& & S_{sL,ab}= \\
\nonumber & & \frac{5}{4} \left( \sum_{i,j \neq i}
\frac{(1-\alpha_i)(1-\alpha_j)}{2\alpha_i-1} \pi G_{ab}({\bf z_i},{\bf
z_j},\tau) - 2(\lambda_0 - \ln \left| 2\pi \eta^2(\tau)  \right| ) \sum_i
\frac{(1-\alpha_i)^2}{2\alpha_i-1} + \sum_i f(\alpha_i)
\right) .
\end{eqnarray}

For the odd spin structure the Regge superconformal factor takes the
form 
\begin{equation}
\sum_i (\alpha_i-1) \pi G_{++} ({\bf z},{\bf z_i} ,\tau,\chi)
\end{equation}
with $G_{++}$ given by eq.(\ref{oddgreenfunc}), to which the zero
modes of $\tilde {\Box}^{(-)}$ (see eq. (\ref{oddderiv})) have to be
added. These are given by 
\begin{equation}
\lambda_0 + (c_1\chi +\bar c_2 \bar\chi) \theta -(\bar c_1 \bar\chi +
c_2 \chi) \bar \theta 
\end{equation}
with $\lambda_0$ real and $c_1,c_2$ complex bosonic variables.
Performing a modular transformation the zweibein $\tilde E_M^{~A}$
goes over to a new zweibein given by 
\begin{equation}
\tilde {E'}_M^{~A}= \frac{\partial {\bf z}^N}{\partial {\bf z'}^M}
\tilde E_N^{~A} .
\end{equation}

$\tilde {E'}_M^{~A}$ is not of our canonical form (\ref{oddzweibein})
but can be reduced to it (obviously with $\zeta$ and $\theta$ replaced
by $\zeta'$ and $\theta'$ ) through a super Weyl transformation and a
U(1) internal rotation. The new superconformal factor $\Sigma'({\bf z'})$ is
given by
\begin{equation} \label{trasfsigma}
\Sigma'(\hat{\bf z}') =
\Sigma(\hat{\bf z}) +\ln \left|c \tau+d \right| + 
\frac{c\chi \theta}{c \tau +d} -\frac{c\bar\chi\bar\theta}{c \bar\tau+d}.
\end{equation}
Such a transformation is obtained by taking into account that the
supermodular transformations are 
\begin{equation} \label{simplesupermod}
z'=\frac{z}{c\tau +d}; \qquad \theta'=\frac{\theta}{(c\tau +d)^{1/2}}
\end{equation}  
as follows from eqs. (\ref{supermodular1}) and (\ref{hatvar}) for the
$\hat z, \hat \theta$ variables, and using ${\rm sdet}(\tilde E_M^{~A}) =
1+\zeta\theta  -\bar\zeta\bar\theta$, which transforms according to
(\ref{simplesupermod}) and (\ref{supermodular2}).
The modular invariance of $G_{++}$ implies the following
transformations
\begin{eqnarray} \label{oddparamtransf}
\lambda_0' & = & \lambda_0 + \ln \left| \tau c+d \right| \nonumber \\
c'_1 & = & c_1(c\tau +d)^2 + c(c\tau +d) \nonumber \\
c'_2 & = & c_2(c\bar\tau +d)^{\frac{1}{2}}(c\tau +d)^{\frac{3}{2}}.
\end{eqnarray}

By exploiting the independence from the topology of the singular
behaviour of the action at short distances we obtain the structure
\begin{equation}
S_{sL} = {\rm const} \left[ \sum_{i, j \neq i}
\frac{(1-\alpha_i)(1-\alpha_j)}{2\alpha_i-1} \pi G_{++}({\bf z_i},{\bf
z_j},\tau,\chi) + B(\alpha,\lambda_0,\tau,\chi,c_1\chi,c_2\chi) \right] .
\end{equation}
In order to obtain a more explicit form of $B$ we consider first the
case $\chi =0$. $B$ now depends only on $\alpha, \lambda_0,\tau$ and
using the same argument as for the even spin structures we obtain
\begin{equation}
B(\alpha,\lambda_0,\tau,0,0,0) = -2(\lambda_0 - \ln
\left|2\pi\eta^2(\tau) \right| ) \sum_i
\frac{(1-\alpha_i)^2}{2\alpha_i-1} + \sum_i f(\alpha_i).
\end{equation}
Because of the nihilpotency of $\chi$ and the bosonic nature of the
action we now have 
\begin{equation}
B(\alpha,\lambda_0,\tau,\chi,c_1\chi,c_2\chi) =
B(\alpha,\lambda_0,\tau,0,0,0) + \chi \bar\chi \left[ F_0 + \sum_{m,n
=1}^2 F_{mn}c_m \bar c_n + (\sum_{m=1}^2 F_m c_m + c.c.) \right] 
\end{equation}
where the $F$'s are functions of the modular invariant $\alpha_i$ and
of  $\tau$.

Invariance of the action under modular transformations
(\ref{oddparamtransf}) and (\ref{supermodular2}) gives a set of
restrictions on the $F$'s, which however are not sufficient to
determine them completely. Probably for the odd spin structure case a
more profound appeal to the structure of the heat kernel derivation is
needed to pinpoint completely the unknown $F$'s.
Summing up, the result we have achieved for the odd spin structure is
the following
\begin{eqnarray}
& &  S_{sL,++}= \nonumber  \\ 
\nonumber & & \frac{5}{4} \left(\sum_{i,j \neq i}
\frac{(1-\alpha_i)(1-\alpha_j)}{2\alpha_i -1} \pi G_{++} ({\bf z_i},
{\bf z_j}, \tau, \chi) -  
2 (\lambda_0 - \ln \left| 2\pi \eta^2(\tau) \right| ) \sum_i
\frac{(1-\alpha_i)^2}{2\alpha_i -1} \right) \\ 
& &  + \sum_i f(\alpha_i) + \chi \bar\chi \left[  F_0(\alpha,\tau) + \sum_{m,n
=1}^2 F_{mn}(\alpha,\tau)c_m \bar c_n + (\sum_{m=1}^2 F_m(\alpha,\tau)
c_m + c.c.) \right] .  
\end{eqnarray}

For amplitudes to which both even and odd spin structures contribute
the relative weight is fixed by the factorization property for
multiloop amplitudes \cite{cdf}. 
\bigskip
\bigskip

\noindent
{\bf Acknowledgments}

\noindent
We are grateful to Pietro Fr\`e for a useful discussion.

\bibliographystyle{plain}

\end{document}